\documentclass[aps,pra,amssymb,superscriptaddress,10pt,tightenlines]{revtex4}

\usepackage[english]{babel}
\usepackage{graphicx}
\usepackage{dcolumn}
\usepackage{bm}
\usepackage{color}
\usepackage{subfigure}
\usepackage[ansinew]{inputenc}
\usepackage{amsfonts}
\usepackage{amsmath}
\usepackage{amssymb}
\usepackage{subfigure}

\usepackage[braket]{qcircuit}
\usepackage{braket}

\usepackage{mathrsfs}
\usepackage{dsfont}

\def\bra#1{\langle #1 |}
\def\ket#1{| #1 \rangle}

\def\e{\mathrm{e}}
\def\d{\mathrm{d}}
\def\ii{\mathrm{i}}

\begin{document}



\begin{center}
\title*{\Large{Dynamical quantum phase transitions of the Schwinger model: \\ real-time dynamics on IBM Quantum}}
\end{center}

\author{Domenico Pomarico}%
\email{domenico.pomarico@ba.infn.it}
\affiliation{%
	Dipartimento di Fisica, Universit\`a di Bari, I-70126 Bari, Italy
}%
\affiliation{
Istituto Nazionale di Fisica Nucleare, Sezione di Bari, I-70126 Bari, Italy}

\author{Leonardo Cosmai}%
\affiliation{
	Istituto Nazionale di Fisica Nucleare, Sezione di Bari, I-70126 Bari, Italy}

\author{Paolo Facchi}%
\affiliation{%
	Dipartimento di Fisica, Universit\`a di Bari, I-70126 Bari, Italy
}%
\affiliation{
	Istituto Nazionale di Fisica Nucleare, Sezione di Bari, I-70126 Bari, Italy}

\author{Cosmo Lupo}%
\affiliation{
	Istituto Nazionale di Fisica Nucleare, Sezione di Bari, I-70126 Bari, Italy}
\affiliation{%
	Dipartimento  di Fisica, Politecnico di Bari, I-70126 Bari, Italy
}%

\author{Saverio Pascazio}%
\affiliation{%
	Dipartimento  di Fisica, Universit\`a di Bari, I-70126 Bari, Italy
}%
\affiliation{
	Istituto Nazionale di Fisica Nucleare, Sezione di Bari, I-70126 Bari, Italy}

\author{Francesco V. Pepe}%
\affiliation{%
	Dipartimento di Fisica, Universit\`a di Bari, I-70126 Bari, Italy
}%
\affiliation{
	Istituto Nazionale di Fisica Nucleare, Sezione di Bari, I-70126 Bari, Italy}

\begin{abstract}
	Simulating real-time dynamics of gauge theories represents a paradigmatic use case to test the hardware capabilities of a quantum computer, since it can involve non-trivial input states preparation, discretized time evolution, long-distance entanglement, and measurement in a noisy environment. We implement an algorithm to simulate the real-time dynamics of a few-qubit system that approximates the Schwinger model in the framework of lattice gauge theories, with specific attention to the occurrence of a dynamical quantum phase transition. Limitations in the simulation capabilities on IBM Quantum are imposed by noise affecting the application of single-qubit and two-qubit gates, which combine in the decomposition of Trotter evolution. The experimental results collected in quantum algorithm runs on IBM Quantum are compared with noise models to characterize the performance in the absence of error mitigation.
\end{abstract}

\maketitle

\section{Introduction}

The availability of noisy intermediate-scale quantum (NISQ) devices in cloud access platforms is a fundamental step towards the quantum computing era. Nonetheless, the limited number of available qubits and the absence of controllable errors probabilities prevents those systems from actually outperforming current classical computing capabilities in any task. Multiple hardware setups have been engineered for a quantum computing purpose, with different advantages regarding gates fidelity and experimental realization. Examples of NISQ devices are represented by circuits with superconducting transmon qubits~\cite{ibmq, cirq, amazon}, ion traps or optical lattices hosting Rydberg atoms~\cite{ionq,fidelity,ionq2,ionq3,ionq4}, qubits encoded in photonic modes of optical setups~\cite{china, xanadu} to cite a few of them.

In this framework, high energy physics represents an interesting testbed for quantum devices. On one hand, quantum computation can be applied ``downstream'', to optimize data analysis from and event reconstruction from experiments~\cite{quanthep1,quanthep2,quanthep3,quanthep4,quanthep5}. On the other hand, the ``upstream'' investigation of gauge theories, especially in their lattice formulations~\cite{lgt1,lgt2,lgt3,lgt4,lgt5}, can benefit from the possibility to perform quantum simulations of regimes not achievable with perturbative techniques. Long-standing questions related to low-energy processes in quantum chromodynamics (QCD) are still far from current capabilities of Monte Carlo techniques, due to the sign problem intrinsically related to fermionic amplitudes~\cite{mc1,mc2,mc3}. To overcome limitations, research at the interface among quantum information, condensed-matter and high-energy physics is targeting the adoption of new theoretical and computational tools. The state of the art in the field is represented by tensor network methods, able to reduce the exponential complexity to a polynomial one for states characterized by short-range entanglement~\cite{tns1,tns2,tns3,tns4,tns5,tns51,tns52}. These methods are suitable candidate to obtain a breakthrough in non-perturbative regimes: some preliminary studies about quantum electrodynamics (QED) in one spatial dimension proved the ability of tensor networks in describing a wide phenomenology, such as vacuum phase transition, string breaking mechanism and scattering processes~\cite{Ising,tns6,tns7}. On the other hand, highly entangled quantum systems must be studied by means of specifically designed setups, since their complexity cannot be managed in current classical computing capabilities. 

A major role in making quantum computation and simulation effective to solve practical problems in NISQ devices is played by error correction and mitigation~\cite{err1,err2,err3}. In the context of digital real-time evolution ~\cite{ibm_qed,dqpt_ibm,plaquette,index} such procedures should keep the quantum state of the system in the physical subspaces allowed by the gauge constraint. Analog simulations in optical lattices can adopt a periodic drive to obtain energy terms endowed with the same symmetry characterizing lattice QED~\cite{exp1,exp2}. In this case, errors can induce gauge-invariance breaking terms, that can lead to an emergent prethermal behavior~\cite{preth1,preth2,preth3,preth4}.

This paper is aimed at testing the superconducting qubit systems available in the IBM Quantum platform~\cite{ibmq} in a simple lattice gauge theory application. We implement digital evolutions generated by the QED Hamiltonian in $1+1$ dimensions, consisting in non-commuting local contributions~\cite{ionq, Zn, Ising, tns6, tns7, QLM, QLMopen, trott, ibmq_lgt, ibm_qed, Ising2, notarnicola2020} and showing the occurrence of dynamical quantum phase transitions (DQPTs)~\cite{dqpt} in specific cases of quantum quenches. The study of real-time dynamics involves three crucial stages: the preparation of initial states, the evolution, and the final measurements. Gates error affecting a digital simulation accumulate in a more or less coherent manner, which is affected by more variables and more error sources than the stand-alone characterization of gates. Time evolution is partitioned into steps to monitor the measurements statistics variation without the inclusion of error correction and mitigation. The ground state preparation required by the chosen quench protocol is specifically designed, in order to minimize errors and characterize the first stage output statistics. Then, we characterize the effectiveness of time evolution, choosing to analyze the system in proximity of a DQPT, where the dynamics is particularly sensitive to noise~\cite{QLMopen}, thus framing the simulation in most unsafe conditions. A concluding estimation of the amount of  error probability reduction required for a partial observation of the targeted DQPT is made by analyzing the statistics of collected results.

The paper is structured as follows. In Section~\ref{schwin}, we introduce the  lattice Schwinger model and the Jordan-Wigner transformation that maps it into a qubit system, and describe the quench protocol and DQPTs expected in the model. In Section~\ref{sim_res}, we describe the experimental scheme composed of ground state preparation and the subsequent Trotter evolution. The collected results are compared with simulated evolution affected by error probabilities of noisy gates. In Section~\ref{discuss}, we relate our results with previous literature and present a possible outlook of our research.

\section{The lattice Schwinger model}\label{schwin}

QED in $1+1$ dimensions, also known as the Schwinger model, is a $\mathrm{U}(1)$ gauge theory describing the interaction of the electromagnetic field, consisting of only an electric component, and a fermionic particle with mass $m$ and charge $g$. The model can be discretized on a one-dimensional lattice with spacing $a$, by associating to each lattice site $x$ an anticommuting field $\psi_x$, which represents a spinless fermion, while links between each pair of neighboring sites host the gauge degrees of freedom, described by the electric field $E_{x,x+1}$ and the vector potential $A_{x,x+1}$. The latter determines the gauge connection $U_{x,x+1} = \e^{\ii a A_{x,x+1}}$, characterized by the generalized canonical commutation relation $[E_{x',x'+1},U_{x,x+1}] = \delta_{x,x'} U_{x,x+1}$. The lattice model Hamiltonian for a finite lattice with $N$ sites reads~\cite{Zn,Ising,tns6,tns7,ionq} 
\begin{equation} \label{Hamiltonian}
H = - \frac{\ii}{2a} \sum_{x=0}^{N-1} \left( \psi_x^\dagger U_{x,x+1} \psi_{x+1} - \mbox{H.c.} \right) + m \sum_{x=0}^{N-1} (-1)^x \psi_x^\dagger \psi_x + \frac{g^2 a}{2} \sum_{x=0}^{N-1} E_{x,x+1}^2,
\end{equation}
where periodic boundary conditions~\cite{QLM,QLMopen} require the  identification $N\equiv 0$. The model involves staggered (Kogut-Susskind) fermions~\cite{stagg}, described by single-component spinors $\psi_x$, with negative-mass components encoded in odd-$x$ sites. 
The physical  subspace $\mathscr{H}_G$ is spanned by states $\ket{\phi}$ satisfying the Gauss law constraint $G_x \ket{\phi} = 0$ at all sites $x$, where, for a $\mathbb{Z}_n$ gauge group,
\begin{equation}
G_x = \sqrt{\frac{n}{2\pi}} \left(E_{x,x+1} - E_{x-1,x}\right) - \psi_x^\dagger \psi_x - \frac{(-1)^x - 1}{2}.
\end{equation}
The electric field is simulated in the following through a $\mathbb{Z}_n$ discretization of U$(1)$~\cite{Zn, Ising, trott, tns6, tns7, notarnicola2020} with $n=2$. Unlike in the quantum link models~\cite{QLM, QLMopen}, where the electric field is replaced by a spin operator, the $\mathbb{Z}_n$ model is based on replacing gauge connections with permutation matrices~\cite{Zn}. In the case of $\mathbb{Z}_2$, the electric field in each link can have two states, that will be labelled as $\{ \ket{\uparrow}, \ket{\downarrow} \}$, with $E_{x,x+1} \ket{\uparrow} = \frac{\sqrt{\pi}}{2} \ket{\uparrow}$ and $E_{x,x+1} \ket{\downarrow} = -\frac{\sqrt{\pi}}{2} \ket{\downarrow}$. The gauge connections act as $U_{x,x+1} \ket{\downarrow} = \ket{\uparrow}$ and $U_{x,x+1} \ket{\uparrow} = \ket{\downarrow}$~\cite{Ising,tns6,tns7}. An immediate implication of the $\mathbb{Z}_2$ model is the irrelevance in the Hamiltonian \eqref{Hamiltonian} of the electric field energy, which becomes a constant.

\begin{figure}[t!]
	\centering
	\subfigure[]{\includegraphics[width=0.24\linewidth]{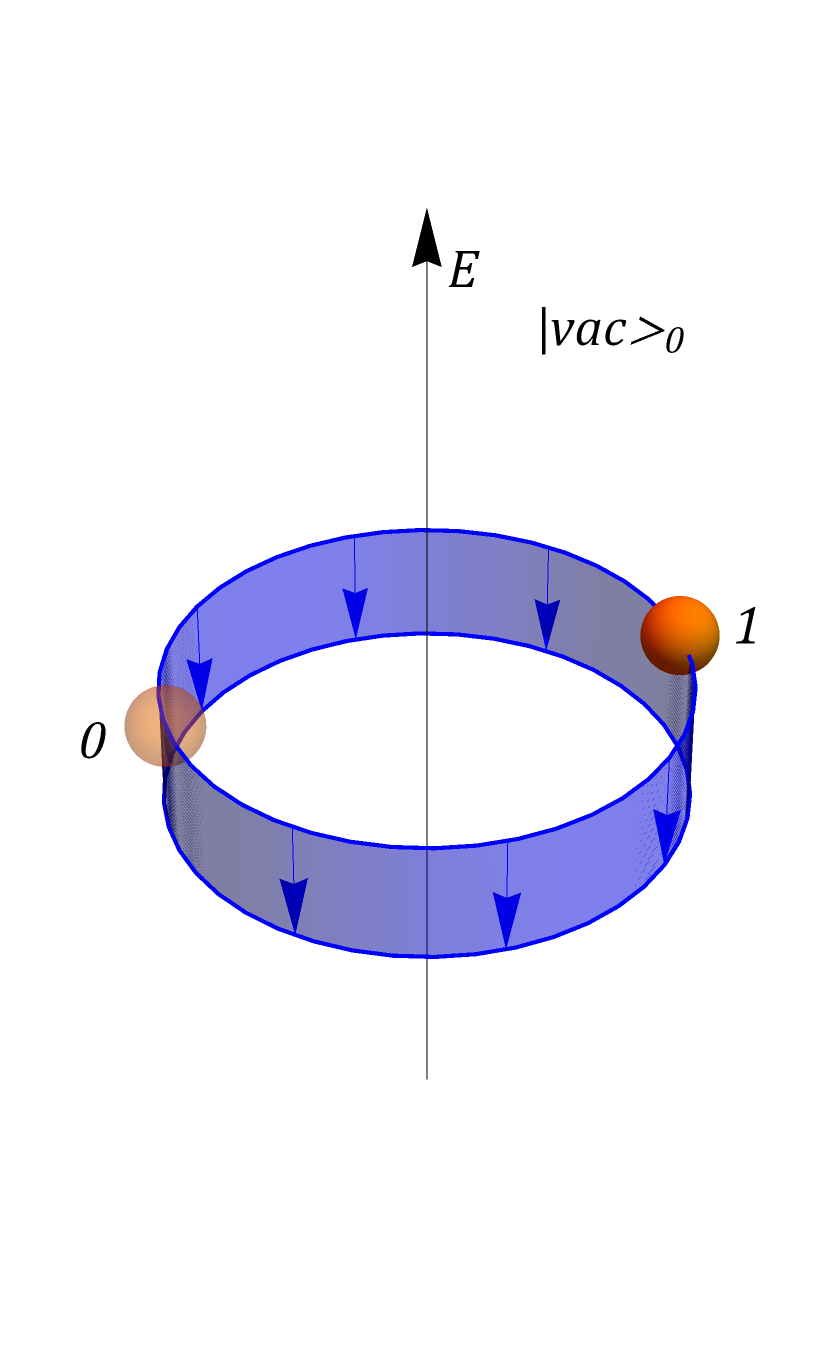}}
	\subfigure[]{\includegraphics[width=0.24\linewidth]{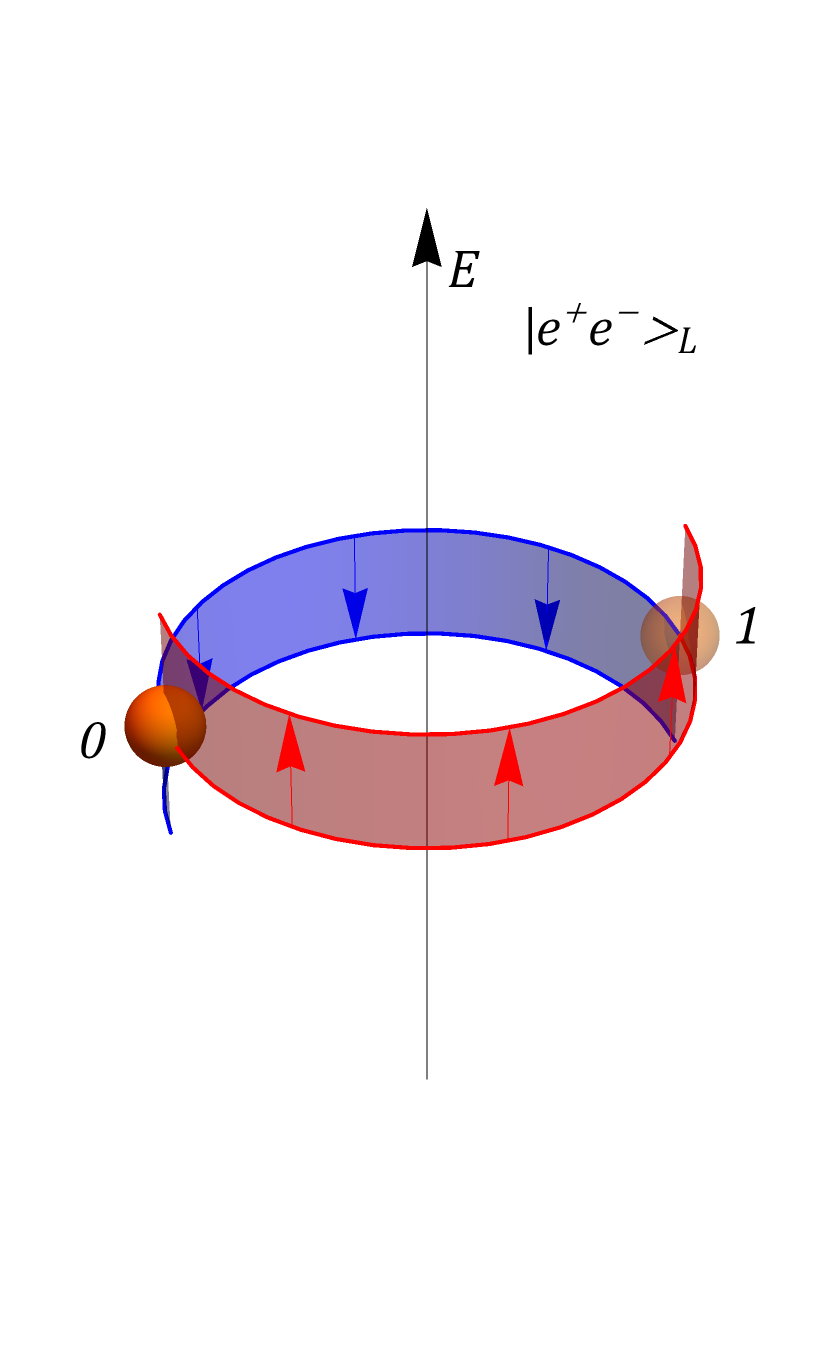}}
	\subfigure[]{\includegraphics[width=0.24\linewidth]{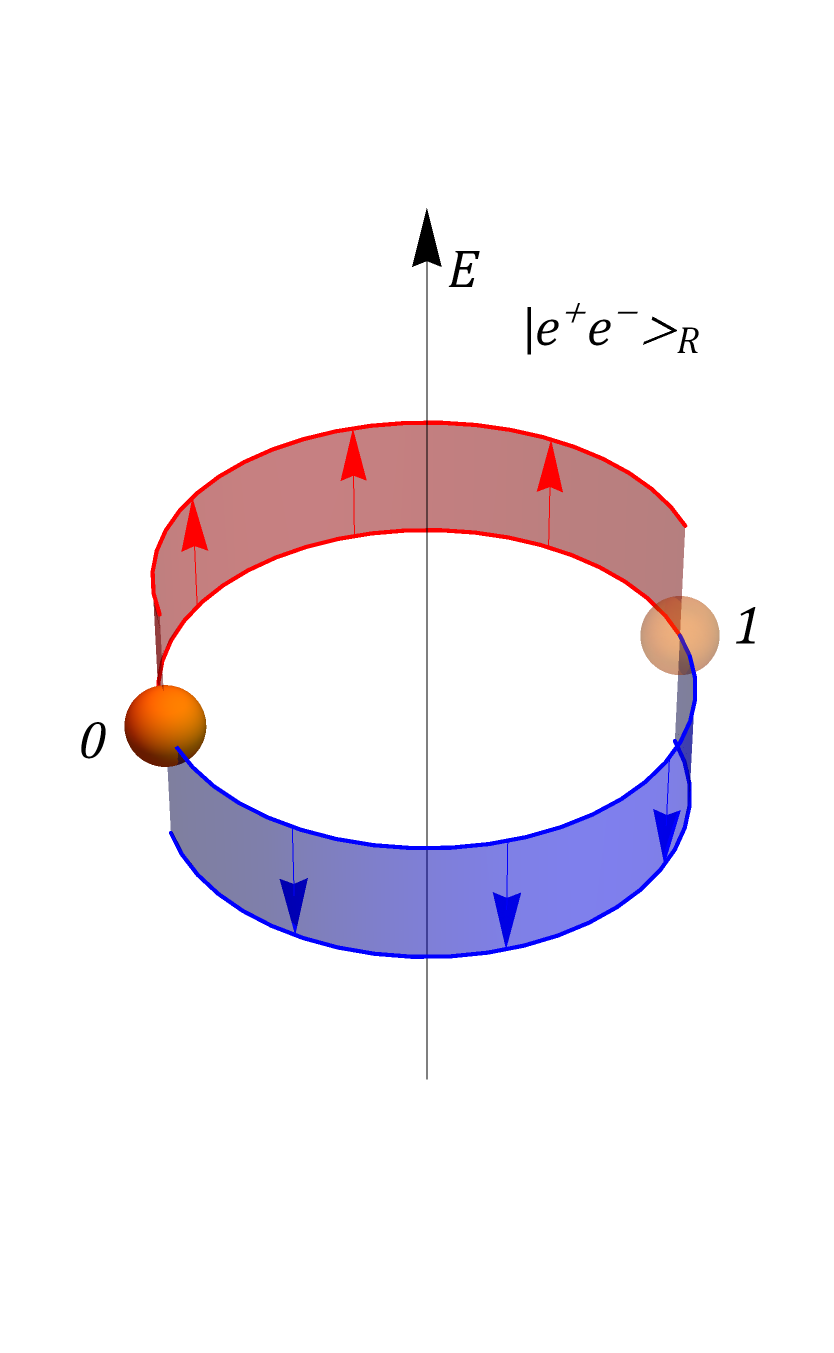}}
	\subfigure[]{\includegraphics[width=0.24\linewidth]{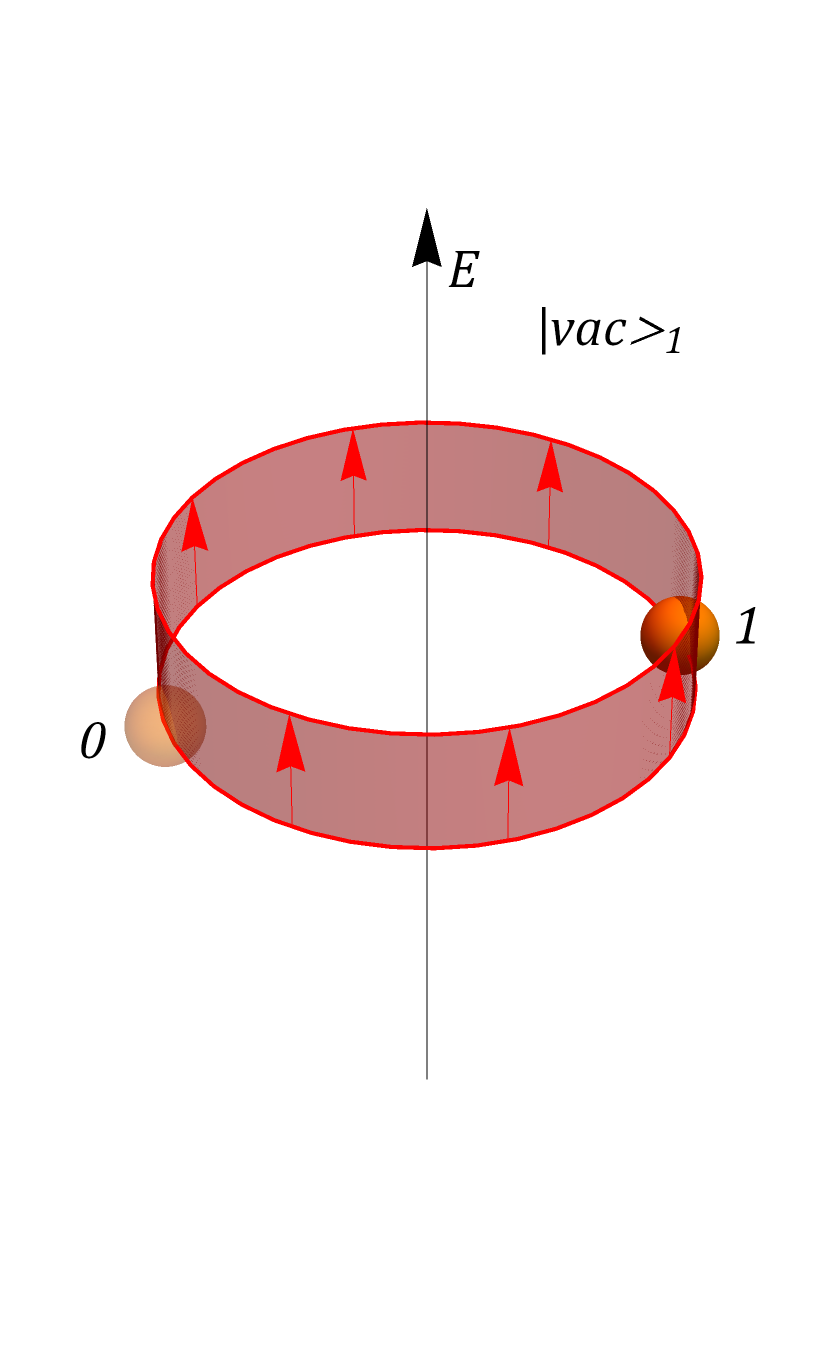}}
	\caption{Representation of physical subspace basis states in a $\mathbb{Z}_n$ gauge model, implemented on a two-site lattice. In all panels, full (transparent) spheres represent occupied (empty) matter sites, while red (blue) edges correspond to a positive (negative) electric field on a link. Panels (a) and (d) represent ``Dirac vacua'', characterized by an occupied negative-mass fermion site, an empty positive-mass fermion site, and a constant background electric field. Panels (b) and (c) represent ``meson'' states, with an occupied positive-mass fermion site, an empty negative-mass fermion site (corresponding to an antiparticle), and a staggered electric field. } \label{phys_sub}
\end{figure}

The simplest nontrivial periodic lattice is composed by $N=2$ sites: the states spanning the physical subspace $\mathscr{H}_G$ of the $\mathbb{Z}_2$ model in this simple case are shown in Fig.~\ref{phys_sub}. The states in panels (a) and (d) represent two ``Dirac vacua'', with a filled negative-mass and an empty positive-mass site. In these states, the total electric field is constant and nonvanishing. These observations motivates the notation $\ket{vac}_{\pm}$ for these two states, where the index is related to the sign of the background electric field. Particle hopping leads to the remaining  ``mesonic'' basis states $\ket{e^+ e^-}_L$ and $\ket{e^+ e^-}_R$, represented in panels (b) and (c), respectively, where the index is referred to the counterclockwise (L) or clockwise (R) hopping of the fermion from the negative- to the positive-mass site.

The Jordan-Wigner transformation maps the spinor field into a spin system~\cite{ionq,trott}, which corresponds to our qubit register, as
\begin{equation}
\psi_x = \prod_{\ell < x}(\ii Z_\ell) \frac{X_x + \ii Y_x}{2}, \qquad \psi_x^\dagger = \prod_{\ell < x}(-\ii Z_\ell) \frac{X_x - \ii Y_x}{2},
\end{equation}
where $X$, $Y$, $Z$ are Pauli matrices, $\sigma^{\pm}=\frac{X \pm \ii Y}{2}$ and occupied sites correspond to qubit states $\ket{\!\downarrow}$. The Hamiltonian of the resulting spin system is
\begin{equation} \label{spin_Ham}
H = H_J + H_m = \frac{J}{2} \sum_{x=0}^{N-1} \left( \sigma_x^- U_{x,x+1} \sigma_{x+1}^+ + \mbox{H.c.} \right) - \frac{m}{2} \sum_{x=0}^{N-1} (-1)^x Z_x = \sum_{x=0}^{N-1} h_x,
\end{equation}
where the free parameter corresponds to a coupling constant $J=\frac{1}{a}$, once energy is scaled in units of mass $m$.

\subsection{Dynamical quantum phase transitions}\label{dqpt_prot}

We aim at studying the non-equilibrium dynamics of the described lattice Schwinger model following a quantum quench~\cite{QLM, QLMopen, dqpt}. Generally, in this protocol, one considers a family of Hamiltonians $H(\gamma)$ that depends on a tunable parameter, and prepares an initial state coinciding with the ground state $\ket{\psi_g}$ of $H_0 = H(\gamma_0)$. At $t=0$, the Hamiltonian suddenly switches to $H = H(\gamma_f)$, determining the evolution $\ket{\psi(t)}=\e^{-\ii t H} \ket{\psi_g}$, characterized by the survival (or Loschmidt) amplitude 
\begin{equation}
\mathcal{G}(t) = \braket{\psi_g | \psi(t)}.
\end{equation}
To identify a possible DQFT, we search for the zeros of the \textit{Loschmidt echo} 
\begin{equation}
\mathcal{L}(t) = |\mathcal{G}(t)|^2 = \e^{-n \lambda(t)}, 
\end{equation}
which depends on the number $n$ of degrees of freedom and on the rate function $\lambda(t)$, which becomes divergent in correspondence of the aforementioned zeros.

\begin{figure}[t!]
		\centering
		\subfigure[]{\includegraphics[width=0.25\linewidth]{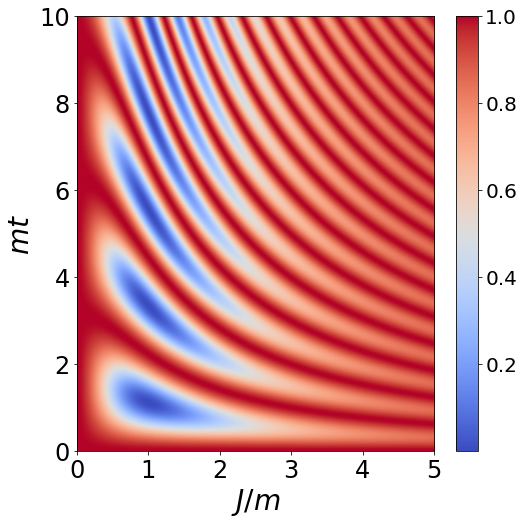}}
		\subfigure[]{\includegraphics[width=0.24\linewidth]{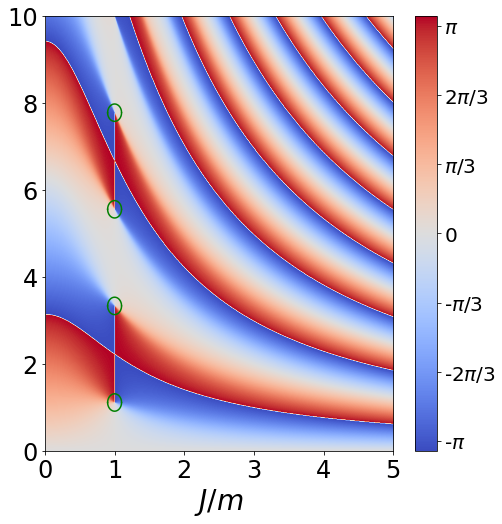}}
		\subfigure[]{\includegraphics[width=0.45\linewidth]{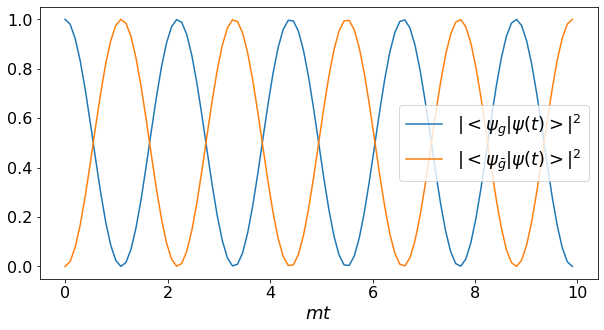}}
	\caption{Panel (a) shows the Loschmidt echo $\mathcal{L}(t)=|\braket{\psi_g | \psi(t)}|^2$ with the variation of the free parameter $J$, while the phase of Loschmidt amplitude is represented in panel (b); here, the green paths around the DQPT points are characterized by nonvanishing winding number. The Trotter evolution discussed in Section~\ref{trott_sec} of Rabi states with step $\Delta t = 0.1$, corresponding to $J/m=1$ and without noise, is shown in panel (c).} \label{rabi_pict}
\end{figure}

The targeted evolution generated by the Hamiltonian \eqref{spin_Ham} is determined by a single free parameter, as the Loschmidt amplitude phase $\varphi(J,t) = \mbox{arg} \ \mathcal{G}(t)$ is undefined in correspondence of the critical points. Nonetheless in their neighbourhood $\varphi$ is expected to be smooth up to a discontinuity line of $2 \pi$ starting from the criticality. This characterization corresponds to a vortex, with a winding number
\begin{equation}
\nu = \frac{1}{2 \pi} \oint_{\mathcal{C}} \mathrm{d} \bm{s} \cdot \bm{\nabla} \varphi,
\end{equation}
where $\mathcal{C}$ is a loop in the $(J,t)$ plane~\cite{QLM}.

The adopted protocol quenches the Kogut-Susskind staggered fermions at $t=0$ by inverting the mass sign: $H(m,J) \longrightarrow H(-m,J)$. The quenched Hamiltonian can be decomposed into parity sectors, as described in Appendix~\ref{appA},
\begin{equation}
H(-m,J) = H^{(-)} \oplus H^{(+)} .
\end{equation}
In the even sector, the evolution is generated by
\begin{equation} \label{rabi}
H^{(+)} = \begin{pmatrix}
-\frac{J^2-m^2}{\sqrt{m^2 + J^2}} & \frac{2Jm}{\sqrt{m^2 + J^2}} \\
\frac{2Jm}{\sqrt{m^2 + J^2}} & \frac{J^2-m^2}{\sqrt{m^2 + J^2}} 
\end{pmatrix},
\end{equation}
in the subspace spanned by the basis $\{\ket{\psi_g},\ket{\psi_{\bar{g}}}\}$, made of the even eigenstates of the initial Hamiltonian $H_0 = H(m,J)$, which are associated to the lowest and highest eigenvalue $E_{g} = - \sqrt{m^2 + J^2}$ and $E_{\bar{g}} = + \sqrt{m^2+J^2}$, respectively. The odd parity sector involves the eigenstates $\ket{\psi_e}$ and $\ket{\psi_{\bar{e}}}$ of $H(m,J)$, which are independent of $J$ and characterized by the eigenvalues $\pm m$; these two states are still eigenstates of the quenched Hamiltonian $H(-m,J)$, which only inverts their eigenvalues.

The Loschmidt amplitude for the initial ground state 
\begin{equation}
\ket{\psi_g} = a_g (\ket{vac}_+ + \ket{vac}_-) + b_g (\ket{e^+ e^-}_L + \ket{e^+ e^-}_R) ,
\end{equation}
derived in Appendix~\ref{appA}, reads
\begin{equation}
\mathcal{G}(t) = (2 a_g^2 - 2 b_g^2)^2 \e^{-\ii E_g t} \left( 1+ \frac{J^2}{m^2} \e^{-\ii (E_{\bar{g}}-E_{g})t} \right),
\end{equation}
with
\begin{equation}
a_{g}=\frac{1}{\sqrt{2 \left(1+p^2_{g}\right)}}, \quad  b_{g}=\frac{p_g}{\sqrt{2 \left(1+p^2_{g}\right)}}, \quad \text{with \,} p_{g}=\frac{m}{J} - \sqrt{\frac{m^2}{J^2}+1} .
\end{equation}
DQPTs are observed for $J = m$ at times 
\begin{equation}
t_j = \frac{(2j+1)\pi}{2 E_{\bar{g}}} = \frac{(2j+1)\pi}{2 \sqrt{2} \ m},
\end{equation}
yielding the Rabi oscillations between $\ket{\psi_g}$ and $\ket{\psi_{\bar{g}}}$ expected from Eq.~\eqref{rabi}, as shown in Fig.~\ref{rabi_pict}. The behavior of the phase, reported in Fig.~\ref{rabi_pict}(b), features vortices corresponding to Loschmidt echo nodes, while the remaining discontinuities in survival maximum values compensate each other.

\subsection{Ground state preparation}\label{ground_prep_sec}

The protocol presented in Section~\ref{dqpt_prot} requires the preparation of the input state $\ket{\psi_g}$, namely the ground state of $H(m,J)$. Based on reasons clarified in Section~\ref{trott_sec}, the degrees of freedom of the lattice are assigned to the four qubits of the \texttt{ibmq\_manila} circuit $\ket{q_0 q_1 q_2 q_3}$
\begin{itemize}
	\item $q_0$ and $q_3$ host the ``electric field'' states of the $\mathbb{Z}_2$ links;
	\item the staggered spinless fermions are encoded in $q_1$ and $q_2$.
\end{itemize}
The four physical states are referred to the following computational basis states: $\ket{vac}_-=\ket{1 0 1 1}$, $\ket{e^+ e^-}_L=\ket{0 1 0 1}$, $\ket{e^+ e^-}_R=\ket{1 1 0 0}$, $\ket{vac}_+=\ket{0 0 1 0}$, expressed according to the IBM \texttt{qiskit} notation $\ket{\uparrow}=\ket{0}$ and $\ket{\downarrow}=\ket{1}$. Since each state can be unambiguously identified by the first two qubits $\ket{q_0 q_1}$, one can associate to the ground state $\ket{\psi_g}$ an auxiliary \textit{product} state of two qubits,
\begin{equation}
\ket{\psi_g'} = a_g(\ket{10}+\ket{00}) + b_g(\ket{01}+\ket{11})= \frac{1}{\sqrt{2}}(\ket{0} + \ket{1}) \otimes \sqrt{2}(a_g \ket{0} + b_g \ket{1}),
\end{equation}
with the amplitudes corresponding, in the DQPT condition $J/m=1$, to $a_g = 0.653$ and $b_g = -0.271$.

\begin{figure}[t!]
	\centering
	\includegraphics[width=0.5\linewidth]{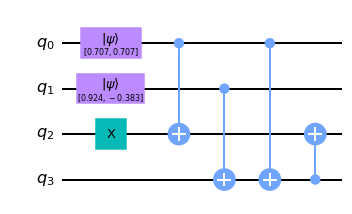}
	\caption{Circuit for the ground state preparation of $H(m,J)$ corresponding to the dynamical quantum phase transition value $J/m=1$. Matter sites correspond to $q_1$ and $q_2$, while $\mathbb{Z}_2$ links are encoded in $q_0$ and $q_3$.} \label{ground_prep}
\end{figure}

The ground state for the complete four-qubit system is obtained by acting with $CNOT$ two-qubit gates
\begin{equation}
CNOT_{ij} \ket{q_i q_j}= \ket{q_i, q_i \oplus q_j},
\end{equation}
which increase the amount of entanglement in the system. For this reason, containing the error probability entailed by these gates is essential to guarantee an effective quantum computation, which cannot be efficiently simulated by classical computers. The circuit chosen for ground state preparation reads
\begin{equation}
\ket{\psi_g} = \ CNOT_{32} \ CNOT_{03} \ CNOT_{13} \ CNOT_{02} \ X_2 \ket{\psi_g'} \otimes \ket{00},
\end{equation}
and is pictorially represented in Fig.~\ref{ground_prep}.

The simulation of the circuits includes an error probability entailed by each gate application, generally described by the  bit-flip and phase-flip error channel $\rho \mapsto \mathcal{D}(\rho)=\sum_{i=0}^3 K_i \rho K_i^\dagger$, with 
\begin{equation} \label{kraus}
K_0 = \sqrt{1-p_x-p_y-p_z} \ \mathds{1}, \quad \ K_1 = \sqrt{p_x} \ X, \quad\ K_2 = \sqrt{p_y} \ Y, \quad \ K_3 = \sqrt{p_z} \ Z.
\end{equation}
Such quantum channels are associated with every single-qubit gate employed in the state preparation, while for the two-qubit gates the independent error probabilities $(p_x, p_y, p_z)$ can be varied in Eq.~\eqref{kraus} to define
\begin{equation} \label{channel}
\rho \mapsto \widetilde{\mathcal{D}}(\rho)=\sum_{i,j=0}^3 \widetilde{K}_{ij} \rho \widetilde{K}_{ij}^\dagger \qquad \mbox{with} \qquad \widetilde{K}_{ij}=K_i \otimes K_j
\end{equation}
for a two-qubit density matrix $\rho$. Each circuit includes also reset and measurement gates, which are affected in simulations only by bit flips~\cite{noise,readout}, implemented by a single noise contribution $K_1$, thus corresponding to $p_z = p_y = 0$.

The comparison of the simulations with the outputs of IBM Quantum is evaluated in terms of the trace distance
\begin{equation} \label{trace}
T(\rho_{\mathrm{ibmq}},\rho_{\mathrm{sim}})=\frac{1}{2} || \rho_{\mathrm{ibmq}} - \rho_{\mathrm{sim}} ||_1 = \frac{1}{2} \mbox{Tr} \left[\sqrt{(\rho_{\mathrm{ibmq}} - \rho_{\mathrm{sim}})^\dagger (\rho_{\mathrm{ibmq}} - \rho_{\mathrm{sim}})} \right] ,
\end{equation}
which quantifies the similarity between the output state of simulated state $\rho_{\mathrm{sim}}$ and the actual output of IBM hardware $\rho_{\mathrm{ibmq}}$.

\subsection{Trotter evolution}\label{trott_sec}

The evolution determined by the Hamiltonian \eqref{spin_Ham}, composed of non-commuting local terms $h_x$, can be approximated by a Trotter decomposition bases on local unitary operators:
\begin{equation} \label{trott_form}
\e^{-\ii H t} = \e^{-\ii \sum_x h_x t} = \left( \e^{-\ii h_{N-1} \Delta t} \e^{-\ii h_{N-2} \Delta t} \dots \e^{-\ii h_0 \Delta t} \right)^{\frac{t}{\Delta t}} + \mathcal{O}(\Delta t) .
\end{equation}
The improved approximation that would in principle be provided by the Suzuki-Trotter formula~\cite{trott} is not well suited in this framework, because it would require a larger number of gates for circuit implementation.

\begin{figure}[t!]
		\includegraphics[width=0.98\linewidth]{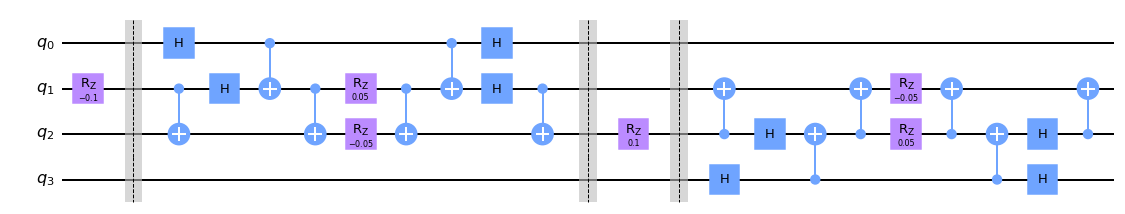}
	\caption{Trotter step as given in Ref.~\cite{trott} for the $\mathbb{Z}_2$ gauge group discretization of lattice QED. Gauge degrees of freedom are encoded in qubits $q_0$ and $q_3$, while fermionic matter is described by qubits $q_1$ and $q_2$. The parameters used in $R_z$ gates correspond to the choice $J=m=1$ and $\Delta t = 0.1$.} \label{trotter}
\end{figure}

A decomposition of each term in Eq.~\eqref{trott_form} according to the available set of gates is formulated in Ref.~\cite{trott}. Here, we present its specific application to the $\mathbb{Z}_2$  gauge group~\cite{Zn, Ising}, where $U_{x,x+1} = U_{x,x+1}^\dagger = X_{x,x+1}$. The fermionic hopping contribution can be equivalently expressed as
\begin{equation} \label{3spin}
H_J = \frac{J}{4} \sum_{x=0}^{N-1} \left[ X_{x,x+1} \left(X_x X_{x+1} + Y_x Y_{x+1} \right) \right] = \sum_{x=0}^{N-1} h_{J,x} .
\end{equation}
The evolution related to the Trotter time steps generated by the three qubits interaction in Eq.~\eqref{3spin} is implemented according to the Cartan decomposition~\cite{KAK1,KAK2,KAK3}. Concerning the periodic lattice with $N=2$ sites, we consider for clarity the hopping term $h_{J,0}$, acting on the subsystem $\ket{q_0 q_1 q_2}$:
\begin{align}
\e^{-\ii h_{J,0} \Delta t} & = K^\dagger A K, \\
K & = CNOT_{12} \ CNOT_{01} \ H_1 \ H_0 \ CNOT_{12}, \\
A & = \mathds{1} \otimes R_z(J \Delta t/2) \otimes R_z(-J \Delta t/2),
\end{align}
where $H_i$ is the Hadamard gate acting on $q_i$ and $R_z(\alpha) = \e^{-\ii Z \alpha/2}$. The remaining term $h_{J,1}$ acts in an analogous way on the subsystem $\ket{q_1 q_2 q_3}$, as represented in Fig.~\ref{trotter}.

The decomposition first rotates product basis states,
\[ \Qcircuit @C=1em @R=.7em {
	\lstick{\ket{q_0}} & \qw & \gate{H} & \qw \\
	\lstick{\ket{q_1}} & \ctrl{1} & \gate{H} & \qw \\
	\lstick{\ket{q_2}} & \targ & \qw & \qw \gategroup{2}{4}{3}{4}{.8em}{\}}
} \qquad \qquad \qquad \qquad \qquad \qquad \qquad \qquad \qquad \qquad \qquad \qquad \qquad \quad \ \ \Qcircuit @C=1em @R=.7em {
	\lstick{H_0 \ \ket{q_0} = \ \ket{\pm} \qquad \qquad \qquad \qquad \qquad \qquad \qquad \qquad \qquad \quad \ \ \ } \\
	\\
	\\
	\\ 
	\lstick{(H_1 \otimes \mathds{1}) \ CNOT_{12} \ket{q_1 q_2} = \frac{\ket{0} + (-1)^{q_1} \ket{1}}{\sqrt{2}} \otimes \ket{q_1 \oplus q_2} = \ket{Q_{q_1 q_2}}} }
\]
with $\ket{\pm}=\ket{(-1)^{q_0}}=\frac{1}{\sqrt{2}} (\ket{0} \pm \ket{1})$ the $X$ eigenstates. The action of Hadamard gate $H_0$ is required to entangle the state of the associated link with matter sites in the following steps
	\[ \Qcircuit @C=1em @R=1.6em {
		\lstick{\ket{\pm}} & \ctrl{1} & \qw & \qw \\
		\lstick{} & \targ & \ctrl{1} & \qw \\
		\lstick{\raisebox{2.5em}{$\ket{Q_{q_1 q_2}}$}} & \qw & \targ & \qw \gategroup{1}{4}{3}{4}{.8em}{\}}
	} \qquad \qquad \qquad \qquad \qquad \qquad \qquad \qquad \qquad \qquad \qquad \qquad \qquad \qquad \qquad \qquad \qquad \qquad \quad \ \ \Qcircuit @C=1em @R=.7em {
		\\
		\lstick{CNOT_{12} \ CNOT_{01} \ \ket{\pm} \otimes \ket{Q_{q_1 q_2}} = CNOT_{12} \ \frac{\ket{00} \pm \ket{11} + (-1)^{q_1}(\ket{01} \pm \ket{10})}{2} \otimes \ket{q_1 \oplus q_2}} \\
		\\
		\\
		\\ 
		\lstick{= \frac{1}{2} \left[\left(\ket{00} \pm (-1)^{q_1} \ket{10} \right) \otimes \ket{q_1 \oplus q_2} \pm \left( \ket{11} \pm (-1)^{q_1} \ket{01} \right) \otimes \ket{\overline{q_1 \oplus q_2}} \right] = \ket{ZZ_{q_1 q_2}} } } 
	\]
where the bar stands for the logical NOT. A further elaboration of the above $\ket{ZZ_{q_1 q_2}}$ states expression simplifies the application of Cartan decomposition in the evolution with diagonal operators
	\begin{equation}
	\begin{split}
	\ket{ZZ_{q_1 q_2}(\Delta t)} = & \ A \ket{ZZ_{q_1 q_2}} = \mathds{1} \otimes R_z(J \Delta t/2) \otimes R_z(-J \Delta t/2) \left[ \ket{(-1)^{q_1} \pm} \otimes \frac{1}{\sqrt{2}} \left( \ket{0, q_1 \oplus q_2} + (-1)^{q_1} \ket{1, \overline{q_1 \oplus q_2}} \right) \right] \\
	= & \ \ket{(-1)^{q_1} \pm} \otimes \frac{1}{\sqrt{2}} \left( \e^{-\ii \frac{J \Delta t}{4} (1-(-1)^{q_1 \oplus q_2})} \ket{0, q_1 \oplus q_2} + (-1)^{q_1} \e^{\ii \frac{J \Delta t}{4} (1+(-1)^{\overline{q_1 \oplus q_2}})} \ket{1, \overline{q_1 \oplus q_2}} \right),
	\end{split}
	\end{equation}
such that states satisfying $q_1 \oplus q_2 = 1$ acquire a time-dependent phase, while there is no evolution outside the physical subspace $\mathscr{H}_G$ with $q_1 \oplus q_2 = 0$, as predicted by Eq.~\eqref{spin_Ham}. 

Focusing on states with $q_1 \oplus q_2 = 1$, the next circuit steps related with $K^\dagger$ yield
\begin{equation}
\begin{split}
\ket{Q_{q_1 q_2}(\Delta t)} = & \ CNOT_{01} CNOT_{12} \ket{ZZ_{q_1 q_2}(\Delta t)} \\
=& \ \frac{1}{2} \left( \e^{-\ii \frac{J \Delta t}{2}} \ket{00} \pm (-1)^{q_1} \e^{-\ii \frac{J \Delta t}{2}} \ket{11} + (-1)^{q_1} \e^{\ii \frac{J \Delta t}{2}} \ket{01} \pm \e^{\ii \frac{J \Delta t}{2}} \ket{10} \right) \otimes \ket{1},
\end{split}
\end{equation}
followed by the last part of the decomposition
\begin{equation}
CNOT_{12} H_1 H_0 \ket{Q_{q_1 q_2}(\Delta t)} = \begin{cases}
\cos \frac{J \Delta t}{2} \ket{001} - \ii \sin \frac{J \Delta t}{2} \ket{110}, \ \mbox{if} \ q_0=0,q_1=0, \\
\cos \frac{J \Delta t}{2} \ket{101} - \ii \sin \frac{J \Delta t}{2} \ket{010}, \ \mbox{if} \ q_0=1,q_1=0, \\
\cos \frac{J \Delta t}{2} \ket{010} - \ii \sin \frac{J \Delta t}{2} \ket{101}, \ \mbox{if} \ q_0=0,q_1=1, \\
\cos \frac{J \Delta t}{2} \ket{110} - \ii \sin \frac{J \Delta t}{2} \ket{001}, \ \mbox{if} \ q_0=1,q_1=1,
\end{cases}
\end{equation}
as expected by the action of $\e^{-\ii H_{J,0} \Delta t}$. Depending on states of the remaining link, encoded in $\ket{q_3}$, there are states not belonging to $\mathscr{H}_G$ which show a time evolution: they correspond to the ones shown in Fig.~\ref{phys_sub} with a reversed matter sites occupation. The remaining contributions in the Trotter expansion of Eq.~\eqref{trott_form} are the diagonal mass terms $H_m = \sum_{x=0}^{N-1} h_{m,x}$ of Eq.~\eqref{spin_Ham}, expressed by 
\begin{equation}
\e^{-\ii h_{m,x} \Delta t} = R_z(- (-1)^x m \Delta t) = \e^{\ii (-1)^x m Z_x \Delta t / 2}, 
\end{equation}
as reported in Fig.~\ref{trotter}.

\begin{figure}[t!]
	\centering
	\begin{tabular}{cc}
		\begin{tabular}{c}
			\subfigure[]{\includegraphics[width=0.45\linewidth]{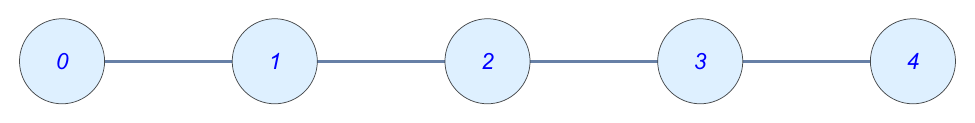}} \\ \subfigure[]{\includegraphics[width=0.25\linewidth]{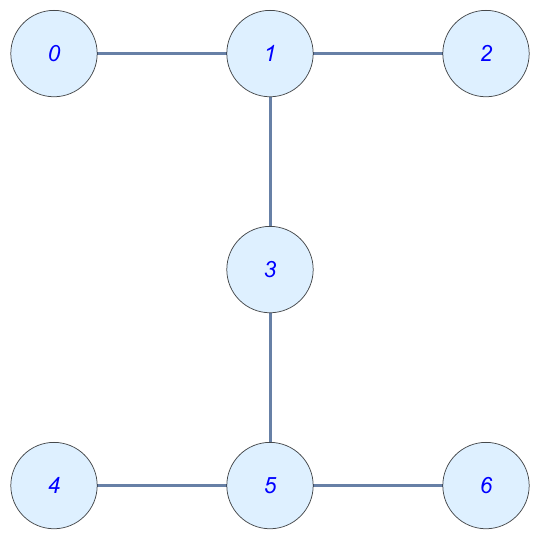}}
			\vspace{5cm}
		\end{tabular} &
		\subfigure[]{\includegraphics[width=0.48\linewidth]{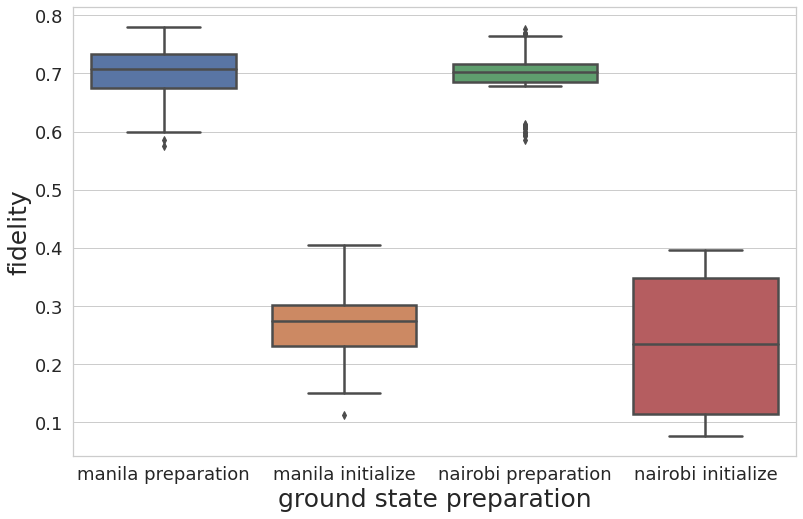}}
	\end{tabular}
	\vspace{-5cm}
	\caption{Topologies of the circuits \texttt{ibmq\_manila}, in panel (a), and \texttt{ibm\_nairobi}, in panel (b). The distributions of ground state preparation fidelity obtained by applying the scheme proposed in Section~\ref{ground_prep_sec} are compared with the output of the built-in command \texttt{QuantumCircuit.initialize} in the boxplot of panel (c).} \label{boxplot}
\end{figure}

The topology of \texttt{ibmq\_manila} in Fig.~\ref{boxplot}(a) is well suited for the implementation of the Trotter evolution, since every CNOT involves nearest-neighbor qubits. In a noiseless scenario, the presented Trotter evolution would yield the Rabi oscillations in Fig.~\ref{rabi_pict}(c), corresponding to the analytical solution of the evolution by the quenched Hamiltonian.

\begin{figure}[t!]
	\centering
	\subfigure[]{\includegraphics[width=0.82\linewidth]{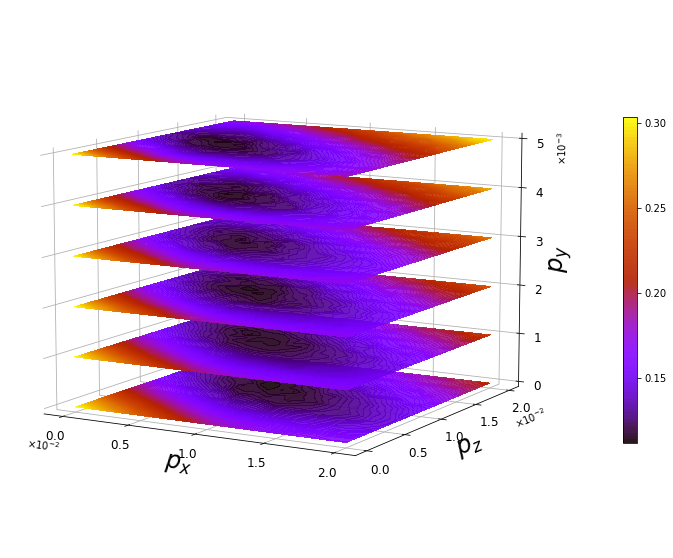}}
	\subfigure[]{\includegraphics[width=0.48\linewidth]{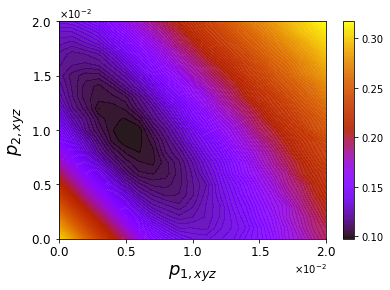}}
	\subfigure[]{\includegraphics[width=0.48\linewidth]{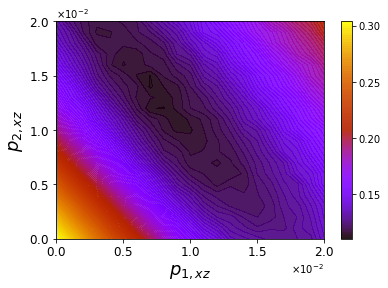}}
	\caption{Trace distance $T(\rho_{\mathrm{ibmq}},\rho_{\mathrm{sim}})$ of the simulated noise models affecting circuits output with respect to the averaged ground state of \texttt{ibmq\_manila} by varying error probabilities. In panel (a) contour plots in the plane $(p_x, p_z)$ are referred to different values of $p_y$. Panels (b-c) evaluate the variation in terms of single and double qubit gate error $(p_1, p_2)$ with the inclusion or exclusion of $Y$ errors respectively.} \label{ground_dist}
\end{figure}

\section{Simulations of real-time dynamics}\label{sim_res}

The experimental results presented in the following are collected in the IBM Quantum platform~\cite{ibmq}. The circuit test is based on the simplest periodic lattice required for the implementation of the Schwinger model described in Section~\ref{schwin}, composed by $N=2$ sites for the matter field and an equal number of links endowed with the $\mathbb{Z}_2$ gauge group. This choice allows for the optimization of the number of gates involved in each Trotter time step, compared to higher-dimensional discretizations of the $\mathrm{U}(1)$ gauge group~\cite{trott}. The simulations include a noise model referred with an error probability affecting each gate, but do not take into account effects related to coherence times, as discussed in Section~\ref{discuss}.

The ground state preparation procedure presented in Section~\ref{ground_prep_sec} performs better in terms of fidelity than the Python package \texttt{qiskit} built-in command \texttt{QuantumCircuit}. \texttt{initialize}, as shown in Fig.~\ref{boxplot}. Errors due to the use of $CNOT$ gates between non-neighboring qubits are investigated by implementing the same preparation in two different topologies, shown in Fig.~\ref{boxplot}(a)-(b): in \texttt{ibm\_nairobi}, the qubits are encoded as follows [see qubit labels in panel (b)]: $q_0 \rightarrow `2$', $q_1 \rightarrow `0$', $q_2 \rightarrow `3$', $q_3 \rightarrow `1$'. Actually, Fig.~\ref{ground_prep} shows the presence of three $CNOT$ gates involving $q_3$, together with each one of the remaining qubits. For this reason, it is convenient to encode it into the highest-degree node of Fig.~\ref{boxplot}(b), while the \texttt{ibm\_nairobi} circuit is limited to the first four qubits. Despite their structural differences, the median values of fidelities obtained with the \texttt{ibm\_manila} and \texttt{ibm\_nairobi} topologies are essentially the same and are about $0.7$, showing in both cases a much higher efficiency with respect to the implementation of \texttt{QuantumCircuit}.\texttt{initialize}. However, the collected statistics in Fig.~\ref{boxplot}(c),  referred to 80 runs for each ground state preparation modality, shows that the interquartile range obtained with \texttt{ibm\_nairobi} is smaller than the one provided by \texttt{ibm\_manila}. Moreover, fluctuations towards low values in the former case are much less relevant.

The readout of the output states is based on \texttt{state\_tomography\_circuits}, that exploits for our 4-qubit circuit the Pauli basis, resulting in $3^4$ circuits required by the related orthogonal measurements~\cite{tomo}. Simulated noise models include the effects of bit flips in the last measurement part of the circuit~\cite{noise, readout}.

The ground state preparation is simulated by defining a noise model in \texttt{AerSimulator}. Three different models are compared with the \texttt{ibmq\_manila} output through the trace distance defined in Eq.~\eqref{trace}~\cite{channel_discr}. Each gate appearing in Fig.~\ref{ground_prep} is affected by error probabilities expressed by the error channels~\eqref{kraus}--\eqref{channel}. The simulations in Fig.~\ref{ground_dist} use the following models of probability assignment:
\begin{itemize}
	\item[(a)] single- and two-qubit gates share the same probability parameters $(p_x,p_y,p_z)$, generally different along the three axes;
	\item[(b)] single-qubit gates have the same error probability along each noise direction $p_1=p_x=p_y=p_z$; two-qubit gates have an analogous property, but are characterized by an independent probability $p_2$;
	\item[(c)] two parameters $p_1$ and $p_2$ quantify the error probability along both $X$ and $Z$ for single- and two-qubit gates, respectively, while errors along $Y$ are neglected.
\end{itemize}

\begin{figure}[t!]
	\centering
		\includegraphics[width=0.95\linewidth]{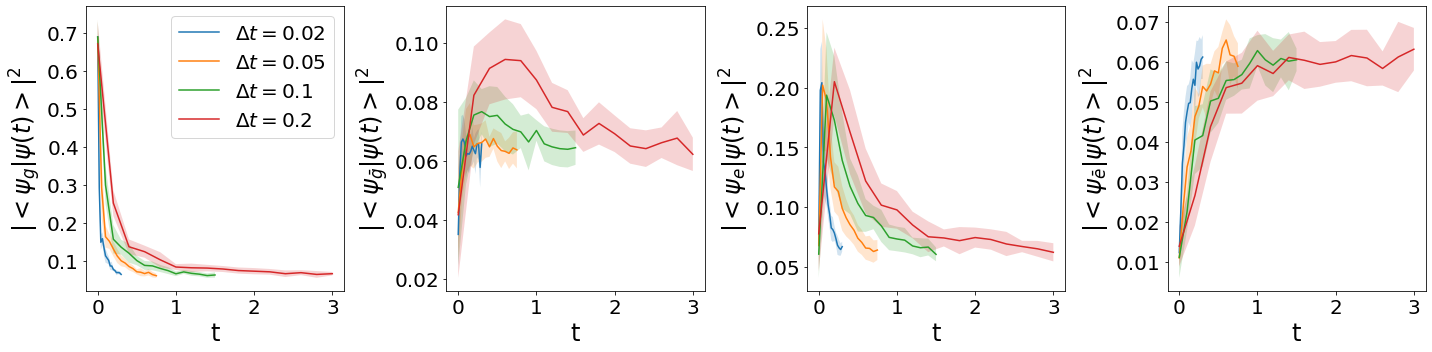}
	\caption{Trotter evolution of the initial state $\ket{\psi_g}$ for different time steps $\Delta t$, run in \texttt{ibmq\_manila}. The average over 10 realizations corresponds to a solid line, while the shaded region represents standard deviation. Time is expressed in units of $m^{-1}$ in all plots.} \label{losch_echo}
\end{figure}

The simulations reported in Fig.~\ref{ground_dist}(a) are averaged over 20 realizations of the noise models, while those in panels (b)--(c) are obtained from 50 realizations. Panel (a) shows a contour plot corresponding to each value of $p_y$, obtained by interpolating the trace distance evaluated over a grid consisting of $21 \times 21$ points with spacing $1 \times 10^{-3}$ in the $(p_x, p_z)$ plane. The considered output of \texttt{ibmq\_manila} is a ground state averaged over 80 realizations. The mean value of the trace distance minimum in the variation of $p_y$ is $10.81 \times 10^{-2} \pm 3.3 \times 10^{-3}$, attained at the averaged coordinates in the plane $(p_x,p_z) = ((10.7 \pm 1.1) \times 10^{-3}, (6.3 \pm 0.9) \times 10^{-3})$. These results are stable with respect to variations in $p_y$. 
Indeed, a $Y$ error is provided by a sequence of simultaneous bit flips and phase flips, whose probability is much smaller than the one of each single error. Panels (b)--(c) show the trace distance of the two noise models with independent parameters for single- and two-qubit gates with respect to the ground state $\rho_{\mathrm{ibmq}}$ experimentally prepared  80 times, with no error along $Y$ in the case of panel $(c)$. The minimum values of the trace distance in the two cases can be considered equal within the statistical fluctuations. 
Moreover, the minima are found in correspondence of $(p_{1,xyz},p_{2,xyz}) \simeq (5 \times 10^{-3},1 \times 10^{-2})$ in panel (b) and $(p_{1,xz},p_{2,xz}) \simeq (7.5 \times 10^{-3},1.5 \times 10^{-2})$ in panel (c), highlighting a scale factor of $3/2$ due to the absence of the error along $Y$ in the latter case. The minimum values of the trace distance, which equal $9.77 \times 10^{-2}$ in panel (b) and $11.16 \times 10^{-2}$ in panel (c), express the ability to distinguish the output $\rho_{\mathrm{sim}}$ from the experimental one $\rho_{\mathrm{ibmq}}$ approximately once out of ten times. 

The ground state in input evolves according to the quench protocol presented in Section~\ref{dqpt_prot}. Non-commuting local terms in the Hamiltonian, determined by the Jordan-Wigner transformation \eqref{spin_Ham}, are circumvented by means of the Trotter evolution described in Section~\ref{trott_sec}, which is implemented using the \texttt{ibmq\_manila} topology. The Loschmidt echo and the overlaps with the remaining physical states are shown in Fig.~\ref{losch_echo}, where the different curves are obtained by varying the time step length $\Delta t$ and by averaging 10 experimental realizations of evolution for each case.

\begin{figure}[b!]
	\centering
	\subfigure[]{\includegraphics[width=0.48\linewidth]{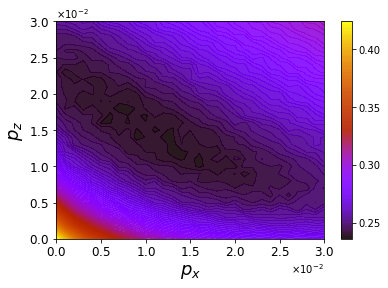}}
	\subfigure[]{\includegraphics[width=0.48\linewidth]{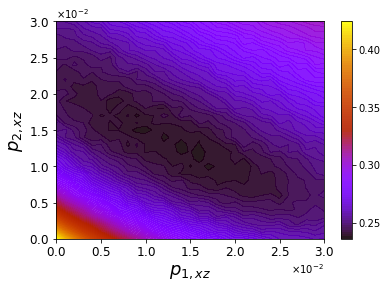}} 
	\caption{Averaged trace distance $\overline{T(\rho_{\mathrm{ibmq}},\rho_{\mathrm{sim}})}$ of the evolution yielded in the first three time steps by \texttt{ibmq\_manila} and the simulated noise models. In panel (a) probability parameters $(p_x, p_z)$ are related with $X$ and $Z$ errors, while, in panel (b), $(p_{1,xz}, p_{2,xz})$ refer to error probabilities in single- and two-qubit gates, respectively, in a case in which $Y$ errors are neglected.} \label{time_dist}
\end{figure}

The high error probability translates into a fast convergence towards the maximally mixed state $\rho_{\infty} = (\mathrm{dim} \ \mathscr{H})^{-1} \mathds{1}$. This trend shows a striking deviation for the probability $|\braket{\psi_{e} | \psi(t)}|^2$ during the first two time steps, probably driven by a coherent error accumulation. To focus on this behavior, as well as to limit the computational resources required in simulations, the trace distance is averaged over the first three time steps $t_i = (i-1) \Delta t$, using
\begin{equation} \label{mean_trace}
\overline{T(\rho_{\mathrm{ibmq}},\rho_{\mathrm{sim}})} = \frac{1}{3 \Delta t} \sum_{i=1}^3 T(\rho_{\mathrm{ibmq}}(t_i),\rho_{\mathrm{sim}}(t_i)) \Delta t,
\end{equation}
expressing the mean probability to distinguish the evolution outputs of \texttt{ibmq\_manila} from the simulated ones.

The optimal error probabilities for the ground state preparation are identified in Fig.~\ref{ground_dist}(c). Simulations of the evolution including noise models fixes these parameters assigned to the gates in Fig.~\ref{ground_prep}. The Trotter evolution given in Fig.~\ref{trotter} includes gates affected by error probabilities, which we implemented according to the following two models:
\begin{itemize}
	\item single- and two-qubit gates are characterized by the same arbitrary parameters $(p_x, p_z)$;
	\item error probabilities in $X$ ad $Z$ directions are equal to each other, but take generally different values for single-qubit gates ($p_1 = p_x = p_z$) and two-qubit gates ($p_2 = p_x = p_z$).
\end{itemize}

\begin{figure}[t!]
	\centering
		\subfigure[]{\includegraphics[width=0.95\linewidth]{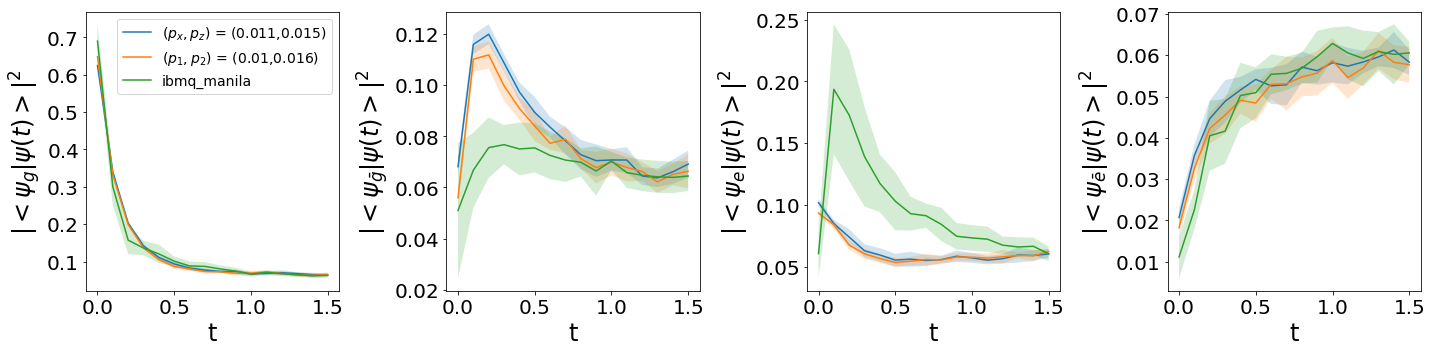}}
		\subfigure[]{\includegraphics[width=0.95\linewidth]{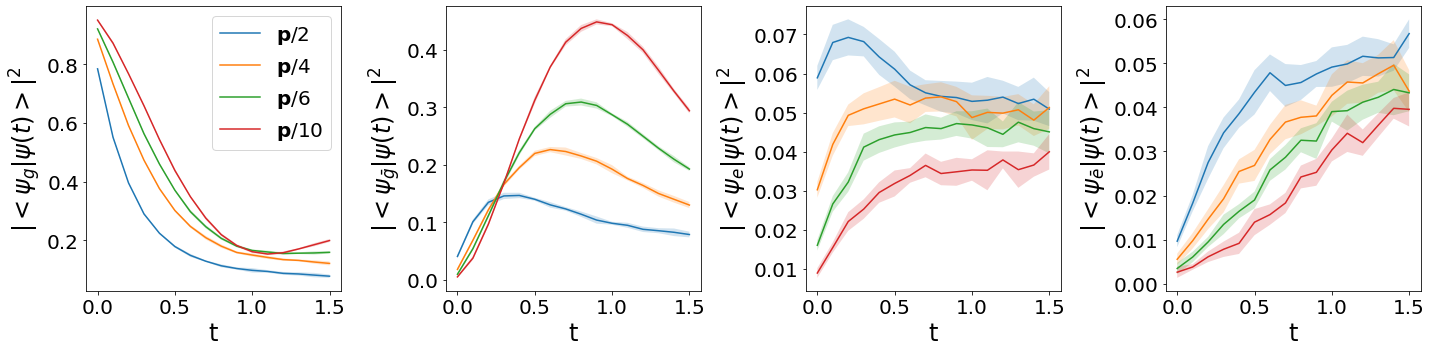}}
	\caption{Noise model implemented in qiskit simulation with Trotter step $\Delta t = 0.1$. In panel (a) we use optimal error probabilities determined in Fig.~\ref{ground_dist}-\ref{time_dist}. In panel (b) such probability array is denoted by $\bf{p}$ to compare different noise regimes in order to observe a revival following the first dynamical quantum phase transition. In all plots, time is expressed in units of $m^{-1}$.}\label{sim}
\end{figure}

The time step that determines the best resolution trade-off in view of investigating the DQPT point is $\Delta t = 0.1$. In the averaged trace distance~\eqref{mean_trace}, the argument $\rho_{\mathrm{ibmq}}$ is averaged over ten experimental realizations of the density matrix evolution, as represented in Fig.~\ref{losch_echo}, while $\rho_{\mathrm{sim}}$ is averaged over 20 noise realizations for each time step of the evolution. The contour plots in Fig.~\ref{time_dist} interpolate the evaluation over a grid composed by $31 \times 31$ points with spacing $10^{-3}$. They describe how much distinguishable both the aforementioned noise models are with respect to the evolution in \texttt{ibmq\_manila}. The minimum value for both models in panel (a)--(b) is obtained by a further evaluation along the elongated direction and it is approximately equal to $23.5 \times 10^{-2}$, corresponding to $(p_x, p_z) \simeq (1.1 \times 10^{-2}, 1.5 \times 10^{-2})$ in the case of panel (a), and $(p_1, p_2) \simeq (1 \times 10^{-2}, 1.6 \times 10^{-2})$ in the case of panel (b). These values are used in Fig.~\ref{sim} to compare the evolution of physical states. In panel (a), the overall behavior characterized by a convergence towards the maximally mixed state is captured by the noise models, but nonnegligible deviations corresponding to the probability $|\braket{\psi_{e} | \psi(t)}|^2$ are still not captured by the model, probably because they are driven by coherently accumulated errors. A comparable high number of single- and two-qubit gates makes the two noise models overlap, as represented in Fig.~\ref{time_dist}, so we can focus on the simpler one described by the probabilities $(p_x,p_z)$. Such a probability array is denoted as $\bf{p}$ in Fig.~\ref{sim}(b), where a comparison of reduced noise regimes aims at estimating a threshold such that a revival is observed after the first DQPT. This translates into the research of a non-monotonic behavior of the Loschmidt echo, which requires an overall error probability per gate ten times lower than in current implementations.

\section{Discussion} \label{discuss}

Error correction and mitigation techniques are expected to limit the quantum advantage due to the required classical information processing. Nevertheless, current simulations on NISQ devices would benefit from the aforementioned techniques in the description of targeted physical phenomena~\cite{err1,err2,err3}. An example is proposed in Ref.~\cite{ibm_qed}, concerning the real-time dynamics on IBM Quantum of a periodic lattice model for a $1+1$ QED model with $N=4$ matter sites. There the discretization of the gauge degrees of freedom is based on a different (non-unitary) truncation. Moreover, the targeted quantities are the vacuum energy and pair production, without a focus on the observation of DQPTs. The exploitation of parity sectors and allowed momenta entails a large reduction of the required qubit number, yielding a scheme able to constrain the evolution in the physical subspace. The reduced Hamiltonian for the matter degrees of freedom in the targeted sector generates a Trotter dynamics implemented in circuits through Cartan decomposition. Aiming at the zero noise extrapolation, a procedure adopting repeated application of noisy $CNOT$ gates is implemented. The circuit depth for each time step increases with respect to the decomposition described in this work, thus allowing us to estimate our targeted first DQPT (with time step $\Delta t = 0.1$) corresponding to the $T_2$ coherence time, because of the 10 time steps limit mentioned in~\cite{ibm_qed}. Indeed, the total number of circuit moments in our Trotter evolution with 10 steps is equal to $200$, as shown in Fig.~\ref{trotter}. We have to include the ground state preparation depth in Fig.~\ref{ground_prep}, because of $SWAP$ gates related with three $CNOT$ between non-nearest neighbors qubits. The maximal gate temporal extent allowed by the coherence time is slightly lower than the effective one, thus signalling an overestimation of our error probabilities.

The comparison of our results with an ion trap simulation of the lattice Schwinger model with $N=4$ matter sites in~\cite{ionq} has to take into account the much lower number of M\o lmer-S\o rensen gates, determined by the higher value of the time step. This is related to the different purpose of the aforementioned work, that aimed at characterizing the pair production mechanism.

The evolution in proximity of a DQPT is considerably more affected by noise~\cite{QLM,QLMopen}, as simulated for a transverse field Ising model~\cite{dqpt_ibm} with error rates comparable with those obtained in our study. Nonetheless, the Hamiltonian terms of an Ising model concern at most spin pairs, thus reducing the circuit complexity for Trotter product formulas. In the case of commuting Hamiltonian terms, at the basis of plaquette dynamics without matter degrees of freedom~\cite{plaquette}, the Trotter product is not required, thus yielding a further reduction of the circuit depth. Concerning the estimated error probabilities in our analysis, their magnitude are confirmed in the study of scalar Yukawa coupling~\cite{yukawa}.

Current experimental realization of ion traps and Rydberg atoms in optical lattices show a higher value for the average gate fidelity of entangling gates~\cite{fidelity,ionq5,rydberg}. The introduction of thermal effects as well as an increased number of parameters for gates errors must be considered in order to improve the proposed noise models. The inclusion of error correction and mitigation~\cite{plaquette,dqpt_ibm,index,ibm_qed} will be investigated in future research to keep the dynamics in the physical subspace and to balance the noise affecting DQPTs observation.

\section{Conclusions}

We study the possibility to simulate real-time dynamics of a model of QED in $1+1$ dimensions, on an elementary lattice composed of two fermionic sites, implemented on IBM Quantum~\cite{ibmq}. More specifically, we analyzed the dynamics after a mass quench, close to a dynamical quantum phase transition. The considered quench protocol requires ground state preparation, based on an optimized circuit able to outperform built-in functions, as measured by fidelity with the ideal state. Limitations in observing DQPTs are described in terms of error probabilities associated to each gate. Different noise models are simulated and compared to capture the main features of the measured evolution, thus determining a marginal contribution of noise along the $Y$ direction. These minimal models reveal the partial observation of the targeted DQPTs phenomena in circuit implementations with a reduced error probability.

\section{Acknowledgements}

D.P. is supported by Regione Puglia and by QuantERA ERA-NET Cofund in Quantum Technologies (Grant No. 731473), project Quantum Computing Solutions for High-Energy Physics (QuantHEP). We acknowledge financial support from PNRR MUR projects PE0000023-NQSTI and CN00000013-National Centre for HPC, Big Data and Quantum Computing. P.F. and S.P. are partially supported by Istituto Nazionale di Fisica Nucleare (INFN) through the project ``QUANTUM''. P.F. is partially supported by the Italian National Group of Mathematical Physics (GNFM-INdAM). We acknowledge the use of IBM Quantum services for this work. The views expressed are those of the authors, and do not reflect the official policy or position of IBM or the IBM Quantum team. Numerical simulations are implemented in ReCaS Bari~\cite{recas}.

\appendix\section[\appendixname~\thesection]{} \label{appA}
The $1+1$ QED Hamiltonian for a periodic lattice with $N=2$ matter sites, analysed in Section~\ref{dqpt_prot} and referred to states in the physical subspace $\mathscr{H}_G$ in Fig.~\ref{phys_sub}, is

\begin{equation} \label{phys_Ham}
\begin{split}
H(m,J) = & \ m \left( \ket{e^+ e^-}_L\bra{e^+ e^-} + \ket{e^+ e^-}_R\bra{e^+ e^-} - \ket{vac}_-\bra{vac} - \ket{vac}_+\bra{vac} \right) \\
& + \frac{J}{2} (\ket{vac}_{- \ L} \bra{e^+ e^-} + \ket{vac}_{- \ R} \bra{e^+ e^-} + \ket{e^+ e^-}_{L \ 1}\bra{vac} + \ket{e^+ e^-}_{R \ 1}\bra{vac} + \mbox{H.c.}),
\end{split}
\end{equation}
in a total Hilbert space with $\mbox{dim} \ \mathscr{H} = 2^{2N}$. By exploiting parity symmetry~\cite{QLM} the eigenstates of the positive sector read
\begin{equation}
\ket{\psi_{g/\bar{g}}} = a_{g/\bar{g}} (\ket{vac}_+ +\ket{vac}_-) + b_{g/\bar{g}} (\ket{e^+ e^-}_L+\ket{e^+ e^-}_R)
\end{equation}
with the coefficients ratio $p_{g/\bar{g}}=\frac{m \mp \sqrt{m^2+J^2}}{J}$ being obtained by imposing $H \ket{\psi_{g/\bar{g}}} = E_{g/\bar{g}} \ket{\psi_{g/\bar{g}}}$. Amplitudes are equal to $a_{g/\bar{g}}=\left[2 \left(1+p^2_{g/\bar{g}}\right)\right]^{-\frac{1}{2}}$ and $b_{g/\bar{g}} = a_{g/\bar{g}} p_{g/\bar{g}}$, thus yielding the Hamiltonian diagonalization with $U = (\ket{\psi_e}, \ket{\psi_{\bar{e}}}, \ket{\psi_g}, \ket{\psi_{\bar{g}}})^\intercal$

\begin{equation} \label{diag_Ham}
U^\dagger H(m,J) U = \mbox{diag}\{ E_e, E_{\bar{e}}, E_g, E_{\bar{g}} \} = \mbox{diag}\left\{ -m, m, -\sqrt{m^2 + J^2}, \sqrt{m^2 + J^2} \right\},
\end{equation}
where the negative parity eigenstates are $\ket{\psi_e}=\frac{1}{\sqrt{2}} (\ket{vac}_+ -\ket{vac}_-)$ and $\ket{\psi_{\bar{e}}}=\frac{1}{\sqrt{2}} (\ket{e^+ e^-}_L-\ket{e^+ e^-}_R)$.The evaluation of expectation values allows us to identify a destructive interference underlying the suppression of hopping contributions in the negative parity sector.

The quench at $t=0$ inverts the mass sign $H(m,J) \longrightarrow H(-m,J)$, whose diagonalization with respect to the previous eigenstates yields a Rabi model in the positive parity sector, while giving just an eigenvalue sign inversion in the negative parity sector. Namely,
\begin{equation}
U^\dagger H(-m,J) U = H^{(-)} \oplus H^{(+)} = \begin{pmatrix}
m & 0 & 0 & 0 \\
0 & -m & 0 & 0 \\
0 & 0 & -\frac{J^2-m^2}{\sqrt{m^2 + J^2}} & \frac{2Jm}{\sqrt{m^2 + J^2}} \\
0 & 0 & \frac{2Jm}{\sqrt{m^2 + J^2}} & \frac{J^2-m^2}{\sqrt{m^2 + J^2}} 
\end{pmatrix},
\end{equation}
with the eigenstate coefficients transformation
\begin{equation}
\begin{split}
b_{\bar{g}} & \longrightarrow b_{\bar{g}}'=-b_{g}, \quad a_{\bar{g}} \longrightarrow a_{\bar{g}}'=a_{g}, \\
b_{g} & \longrightarrow b_{g}'=-b_{\bar{g}}, \quad a_{g} \longrightarrow a_{g}'=a_{\bar{g}}.
\end{split}
\end{equation}

Finally, the analytic expression of the Loschmidt amplitude is obtained by exploiting the completeness relation and reads
\begin{equation}
\begin{split}
\mathcal{G}(t) = & \ \braket{\psi_g | \psi(t)} = \bra{\psi_g} \e^{-\ii H(-m,J) t} \ket{\psi_g} = (2 a_g^2 - 2 b_g^2)^2 \e^{-\ii E_g t} + (2a_g a_{\bar{g}} - 2b_g b_{\bar{g}})^2 \e^{-\ii E_{\bar{g}}t} = \\
= & \ (2 a_g^2 - 2 b_g^2)^2 \e^{-\ii E_g t} \left( 1+ \left( \frac{\widetilde{J}}{\sqrt{1+\widetilde{J}^2}} \left( 1+ \frac{\widetilde{J}^2}{1-\sqrt{1+\widetilde{J}^2}} \right) \right)^2 \e^{-\ii (E_{\bar{g}}-E_{g})t} \right) \\
= & \ (2 a_g^2 - 2 b_g^2)^2 \e^{-\ii E_g t} (1+\widetilde{J}^2 \e^{-\ii (E_{\bar{g}}-E_{g})t} ),
\end{split}
\end{equation}
with a rescaled coupling $\widetilde{J} = J/m$.


\end{document}